\documentstyle[aps,manuscript,epsfig]{revtex}
\begin{document}
\title{Subthreshold K$^+$ production 
in deuteron and alpha induced nuclear reactions}

\author{
M.~D\c{e}bowski$^{1,6}$,
P.~Senger$^1$,
M.~Boivin$^3$,
Y.~Le~Bornec$^2$,
P.~Courtat$^2$,
R.~Gacougnolle$^2$,
E.~Grosse$^{1,4}$,
S.~Kabana$^{1,*}$,
T.~Kirchner$^{4,\dagger}$,
P.~Koczo\'n{}$^1$,
M.~Mang$^1$,
E.~Schwab$^1$,
B.~Tatischeff$^2$,
A.~Wagner$^5$,
W.~Walu\'s{}$^6$,
N.~Willis$^2$,
G.~Wolf$^1$,
R.~Wurzinger$^{3,\ddagger}$,
J.~Yonnet$^3$\\
~\\
$^1${\em Gesellschaft f\"ur Schwerionenforschung, D-64291, Darmstadt, Germany}\\
$^2${\em Institut de Physique Nucl\'{e}aire, IN2P3-CNRS, F-91406 Orsay, France}\\
$^3${\em Laboratoire National Saturne, F-91191 Gif-sur-Yvette Cedex, France}\\
$^4${\em Forschungszentrum Rossendorf, D-01314 Dresden, Germany}\\
$^5${\em Technische Hochschule Darmstadt, D-64289 Darmstadt, Germany}\\
$^6${\em Uniwersytet Jagiello\'{n}ski, PL-30059 Krak\'ow, Poland}\\
} 
\maketitle
\begin{abstract}
Double differential cross sections d$^2\sigma$/dpd$\Omega$ have been
measured for $\pi^+$ and K$^+$ emitted around midrapidity
in d+A and $\alpha$+A collisions at a beam kinetic energy of 1.15~GeV/nucleon.
The total $\pi^+$ yield  increases by a factor of
about 2 when using an alpha projectile instead of a deuteron
whereas the K$^+$ yield increases 
by a factor of about 4. 
According to transport calculations, the K$^+$ enhancement depends 
both on the number of hadron-hadron collisions and on
the energy available in those collisions:
their center-of-mass  energy
increases with increasing number of projectile nucleons.

\end{abstract}
Meson production in nucleus-nucleus 
collisions has become an important tool to study
the dynamics of nuclear matter far from its ground state \cite{stock}.
In particular
K$^+$ mesons are regarded as sensitive probes of the hot and dense reaction zone
due to their long mean free path  \cite{aich,maru,li}.
The sensitivity of kaon production to medium properties is enhanced 
for bombarding energies below and near the K$^+$ threshold
which is 1.58 GeV for free nucleon-nucleon collisions 
(NN$\rightarrow$K$^+\Lambda$N).    
Indeed, it has been found experimentally in Au+Au collisions at 1~GeV/nucleon
that K$^+$ mesons originate predominantly from central collisions
\cite{misko}. 
According to transport calculations, kaons are created  at 
baryonic densities above 2 $\rho_o$ and their enhanced production  
cross section is regarded as an experimental evidence for a  soft nuclear
equation of state \cite{li,hart}.
However, in order to extract information on the properties of dense and hot
baryonic matter from kaon data one has to understand the 
mechanism of subthreshold kaon production and its ingredients:
the elementary processes and their cross sections,
the internal momentum distribution of nucleons,
the role of baryonic resonances
and modifications of the hadron masses in the nuclear medium \cite{fang,broko}. 

It is hardly possible to disentangle experimentally medium effects on kaon 
production by
studying nucleus-nucleus collisions only.
A more transparent situation arises in collisions of protons or very
light projectiles with nuclear targets: in this case 
the K$^+$ mesons are produced in the nuclear medium at normal density.
Up to now angle integrated cross sections for K$^+$ production have been
measured in proton-nucleus collisions at beam energies between
0.8 and 1~GeV \cite{koptev}.
These data can be explained by model calculations which
assume secondary processes $\pi N \rightarrow
K^{+} \Lambda$ to be the most important K$^+$ production channel 
\cite{taras,cass}.
Double differential K$^+$ cross sections have been studied with proton
and deuteron beams on nuclear targets at 2.1~GeV/nucleon \cite{schnetz}.
At this bombarding energy - which is above the K$^+$ threshold - kaons are
produced predominantly in first chance nucleon-nucleon collisions.

In a recent experiment double differential K$^+$ cross sections have been
measured
in proton-nucleus collisions at subthreshold beam energies \cite{debow}.
Only about 10\% of the measured kaon yield can be explained  by first chance
collisions $NN\rightarrow K^{+}YN$ ($Y=\Lambda,\Sigma$).
A similar situation exists
in nucleus-nucleus collisions where transport models have to assume
secondary collisions like $\Delta N\rightarrow K^{+}YN$ in order to reproduce
the measured data \cite{aich,maru,li}.
According to microscopic calculations, multiple hadronic collisions involving
baryonic resonances or pions  are the main source of subthreshold 
kaon production. Such  effects  
can be investigated by experiments which 
use projectiles consisting of  very few nucleons and different nuclear targets.
This Letter reports on a simultaneous measurement of K$^+$ and $\pi^+$
mesons in d and $\alpha$ induced reactions on carbon and lead targets
at a beam energy of 1.15~GeV/nucleon.

The challenge of an experiment on 
subthreshold kaon production in nuclear collisions 
is to handle the huge counting rates  of protons and pions. It requires 
a selective 
and efficient kaon trigger and  techniques to  unambiguously identify the kaons
in spite of the large background produced by rescattered protons. 
The experiment  was performed with 
the magnetic dipole spectrometer SPES~3 at the synchrotron Saturne in Saclay
\cite{debow,comet}. The spectrometer
covered a large solid angle of 10 msr and
a broad momentum range of $p_{max}/p_{min}$=2.3 
(for one magnetic field setting) up to  $p_{max}$=1.4 GeV/c.     
The kaon trigger was based on 
a time-of-flight measurement of momentum selected particles. 
The simultaneous determination of time-of-flight and trajectory
was performed by 4 plastic-scintillator arrays located behind the 
focal plane of the spectrometer. Each of the large-area detector walls 
consisted of 20 vertical paddles. A small plastic scintillator positioned 
in front of the spectrometer provided an additional timing signal. This
detector
allowed a redundant time-of-flight measurement which was used offline 
to effectively suppress the background from rescattered protons.    
Three drift chambers were used for  particle tracking which
results in a momentum 
resolution of $\delta$p/p=10$^{-3}$ and in further rejection 
of background events.
Because of the large flight path of about 10 m from the target 
to the last scintillator wall the counting statistics of low-energy kaons
is limited due to their decay in flight.
The experimental setup and the data analysis have been reported elsewhere
\cite{debow}.

The experiment was performed with beams of d and $\alpha$ particles 
at a  kinetic energy of 1.15~GeV/nucleon and with an intensity of about 
$10^9$ particle/sec. The C and Pb targets had  a thickness 
of 0.68~g/cm$^2$ and 1.35~g/cm$^2$, respectively.
The spectrometer position and the magnetic field were chosen such that 
a polar angular range of $36^{o} < \theta _{lab} < 44^{o}$ and a momentum bite
of 0.35~GeV/c $< p_{lab} <$ 1.15~GeV/c  was covered. These values 
correspond to acceptances of    
0.3 $< y/y_{proj} <$ 0.6  ($70^{o} < \Theta_{cm} < 100^{o}$) for kaons 
and 0.6 $< y/y_{proj} <$ 0.7
($110^{o} < \Theta_{cm} < 120^{o}$) for pions with $y/y_{proj}$ the normalized 
rapidity and $\Theta_{cm}$ the polar angle in the nucleon-nucleon 
center-of-mass system.   
The numbers of kaons registered within a given measuring time  
amount to   
250 K$^+$  in 15 h for $\alpha$+Pb,
100 K$^+$ in 12 h for $\alpha$+C,
130 K$^+$ in 25 h for d+Pb and 
50 K$^+$ in 20 h for d+C.

Figure~1 shows the measured inclusive cross sections $d^{3}\sigma / d p^{3}$
for the production of $\pi^+$ as a function of their kinetic energy in the
nucleon-nucleon center of mass frame (T$_{cm}$).
The statistical errors are smaller than the symbols.
The cross sections have
a systematic error of 18\% due to uncertainties in beam normalization (15\%),
wire chamber efficiency (5\%) and spectrometer acceptance (10\%).
The spectra are fitted by  Maxwell-Boltzmann distributions
$d^{3}\sigma / d p^{3} = A_0 \, exp( - T_{cm}/T_{0})$.
Angle differential cross sections $d\sigma / d\Omega$ are obtained by
integrating the fits over momentum.
The fit parameters and the resulting values for $d\sigma / d\Omega$ 
and $\chi^{2}$ are presented in Table~I.

It is interesting to note that the pion spectra presented in Fig.1 
are ''thermally'' distributed  even for d+C collisions where only  
very few nucleons participate. 
In contrast, pion spectra measured in collisions between
heavy nuclei exhibit an enhancement at low momenta with respect to 
a Maxwell-Boltzmann distribution. This effect was attributed 
to the decay kinematics of the delta resonance 
\cite{brockmann,muentz}.   

The $\pi^+$ cross sections as shown in Fig.1  change 
with the size of the target nucleus and with
the projectile. The target nucleus dependence is demonstrated in the upper part
of Fig.~2 which shows the pion ratio for the two target nuclei
R$_{T}(\pi^{+}) =\frac{d^{3}\sigma}{dp^{3}}(Pb)/\frac{d^{3}\sigma}{dp^{3}}(C)$
as a function of the pion center-of-mass kinetic energy. 
The ratio R$_{T}(\pi^{+})$
increases from 4 to 6
with increasing pion energy, both for the deuteron and the $\alpha$ projectile.
The data obtained with a proton beam are shown for comparison (solid line) 
\cite{debow}.
The assumption that the pion yield is proportional to the surface
of the target nucleus ($\sigma \propto A_{T}^{2/3}$) 
results in a value of  R$_{T} = (\frac{208}{12})^{2/3}$=6.7.
The decrease of the measured ratio R$_{T}(\pi^{+})$ 
towards lower $\pi^+$ energies
indicates an increased absorption of
low momentum pions in Pb as compared to the C nucleus.
The pion production cross section ratio for different projectiles
R$_{P}(\pi^{+}) =
\frac{d^{3}\sigma}{dp^{3}}(\alpha)/\frac{d^{3}\sigma}{dp^{3}}(d)$
(lower part of Fig.~2) increases with increasing pion energy from 1.8 up to
values above 5 for pion kinetic energies of T$_{cm} >$~0.5~GeV.
A value of R$_{P}(\pi^{+})$=2 is expected from
the $\alpha$ to deuteron mass number ratio.
Note, that pions with kinetic energies above T$_{cm} =$~0.32~GeV 
(indicated by an arrow in Fig.2) cannot
be produced in free nucleon-nucleon collision at this bombarding energy.

In the following we will discuss kaon production. In Fig.3 the measured K$^+$ 
cross sections d$^3\sigma$/dp$^3$ are shown as a function of the
kinetic energy in the nucleon-nucleon center-of-mass frame. 
The kaon error bars are due to statistics. The overall systematic
error of 23\% is larger than the one quoted for pions due to the uncertainty
of the kaon identification. The angle-differential K$^+$ production 
cross section d$\sigma$/d$\Omega$ 
is estimated by fitting  a Maxwell-Boltzmann distribution 
to the spectra and integrating over momentum. 
The results of the fits are shown in Fig.3 and the parameters 
are summarized in Table~II.
Because of the poor kaon statistics of the carbon data,
the fit parameters A$_0$ and T$_{0}$ are highly correlated.
Therefore, the carbon data are fitted by varying the amplitude A$_0$ only,
whereas T$_{0}$ is taken from the lead data. This procedure changes the
resulting cross sections by less than 10\% because they are mainly determined
by the first point in the spectrum.

The K$^+$ inverse slope parameters measured with the Pb target are 
significantly lower than the ones  of the corresponding $\pi^+$ spectra.
This indicates the limitation of phase space available in reactions 
involving few nucleons only. In contrast, the K$^+$ spectra measured in 
collisions between heavier nuclei (such as Ne + NaF) at 1 GeV/nucleon
exhibit the same slope as the high energy pions \cite{ahner}. 
 
From Fig.3 and Table~II one can deduce that the K$^+$ yield increases by
a factor of 11.8$\pm$3.3 (for the d projectile)
or a factor of 14.3$\pm$2.5 (for the $\alpha$ projectile)
 when using a Pb target instead of a C target.
This value is larger than the ratio of geometrical cross sections
but is close to the mass ratio of the target nuclei.
The parameterization $\sigma^{K^+} \propto A^{\kappa}_{T}$ gives
$\kappa$=0.87$\pm$0.1 for the deuteron beam and $\kappa$=0.93$\pm$0.06
for the $\alpha$ beam. This result confirms the expectations that K$^+$
are produced all over the reaction volume and are  
not absorbed on their way out of the nucleus in contrast to the pions.
When using an $\alpha$ projectile instead of a deuteron the K$^+$ yield
increases (similarly as the yield of high energy pions) by a factor
\mbox{$\sigma^{K^{+}}(\alpha +A)/\sigma^{K^{+}}(d+A)$=4.3$\pm$0.7} for Pb
and 3.5$\pm$1.0 for the C target.

At beam energies 
above the K$^+$ production threshold a quite different 
dependence of the K$^+$ yield on the projectile mass 
was observed: in d and Ne induced reactions 
at 2.1~GeV/nucleon the K$^+$ cross section 
depends only linearly on
the projectile mass: 
$\sigma^{K^{+}}(Ne+A)/\sigma^{K^{+}}(d+A)$=10$\pm$4 \cite{schnetz}.
In this experiment the K$^+$ yield scales with the size of the
target nucleus according to $\sigma^{K^+} \propto A^{\kappa}_{T}$ with
$\kappa$=0.77$^{+0.17}_{-0.3}$. The errors are due to the uncertainties of 
the K$^+$ production cross sections as quoted in \cite{schnetz}.

In order to understand why the K$^+$ yield increases by nearly a factor of 4 
when the number projectile nucleons increases only by a factor of 2
(as observed at 1.15 GeV/nucleon)  we have performed  
calculations using  a transport equation system of the 
Boltzmann-Uehling-Uhlenbeck type. The details of the model are
given in Ref.\cite{teis}. This code has  
reproduced our data on K$^+$ and $\pi^+$ production  in p+A collisions 
at beam energies of 1.2, 1.5 and 2.5 GeV  \cite{wolf2}. 
We have  calculated the K$^+$ 
double differential cross sections  at $\theta_{lab}$=40$^o$ 
for d+C,Pb and $\alpha$+C,Pb collisions at 1.15 GeV/nucleon.

Figure 4 shows the total K$^+$ yields and their decomposition into the different
contributions according to the BUU model.
The calculations seem to overestimate the K$^+$ data slightly.
We have used the parameterization of the elementary cross sections
as proposed by Zwermann and Sch\"urmann \cite{zwer} and took into
account momentum-dependent nucleon-nucleon interactions. We did not
consider a kaon-nucleon potential which would decrease the kaon yield
\cite{brat}. 
Due to these approximations we do not expect perfect agreement with
the data but rather we want to study the change of the kaon yield
for different target-projectile combinations.   
Both the target and projectile dependence  of the measured K$^+$ 
production cross section is well reproduced by the calculations. 
The total K$^+$ yield 
differs by a factor of about 11 when comparing the BUU results 
for the Pb and the C target and by a factor of about 5 between the d and 
the $\alpha$ projectile. The last factor is found to be practically independent
of the initial internal momentum distribution of the projectiles.

Figure 4 displays 
also the production channels which contribute to the K$^+$ yield.
The most important processes are    
collisions between nucleons and resonances NR$\rightarrow$KYN 
(R = $\Delta_{33}$ ($\approx$40\%) and heavier resonances ($\approx$60\%),
Y = $\Lambda,\Sigma$) and 
between pions and baryons $\pi$B$\rightarrow$KY 
(with B = N ($\approx$90\%) and R ($\approx$10\%)).
The contribution from multiple nucleon-nucleon collisions   
(NN$\rightarrow$KYN) to the kaon yield is only 5-10\%. 
The calculations illustrate that pions and baryonic 
resonances serve as energy reservoirs and thus contribute 
dominantly to subthreshold K$^+$ production.

In the framework of the model calculation the pions are produced via 
decay of baryonic resonances like $\Delta$ and N$^*$. 
The pion data obtained with the Pb target are reasonably well reproduced by the 
calculations whereas for the C target the pion yield is overestimated
by a factor of about 2.
This discrepancy may be due to the fact, that for the light C target 
the pion angular  distribution in the laboratory is strongly forward peaked 
(in contrast to the one for the Pb target). Small errors in
the calculation of the steep rising pion angular distribution
or in  the treatment of pion rescattering change dramatically the pion yield at 
$\theta_{lab}$=40$^o$ for the C target. The total pion yield, however,
which influences the kaon production is not  affected by this 
problem.       

According to our calculations,
the total number of hadron-hadron collisions 
(NR, $\pi$B, NN)  
increases by a factor of 2 when
comparing d+C and $\alpha$+C reactions. In addition, 
the kinetic energies of the projectile nucleons  pile up by multiple
hadron-hadron collisions. This accumulation effect is more efficient for   
larger number of projectile nucleons:
the average center-of-mass energy
of a hadron-hadron encounter which produces a K$^+$ is 
$<{\sqrt s}-{\sqrt s_{th}}>$= 120 MeV in d+C collisions and 
$<{\sqrt s}-{\sqrt s_{th}}>$= 250 MeV in $\alpha$+C collisions
with ${\sqrt s_{th}}$ the K$^+$ threshold energy. 
The difference in the available energies per hadron-hadron collision  
in combination with the steep rise of the elementary K$^+$ excitation functions
near threshold
results in a K$^+$ production cross section which is about 5 times larger
for $\alpha$ induced reactions than for d projectiles.

In summary, we have measured double differential cross sections for
$\pi^{+}$ and K$^+$ mesons emitted in d+C, d+Pb, $\alpha$+C and $\alpha$+Pb
collisions at 1.15~GeV/nucleon.
The cross sections for $\pi^+$ and K$^+$ production
scale approximately with the surface and the mass of the target
nucleus, respectively.
This demonstrates that K$^+$ are produced all over the reaction volume 
and are not absorbed in contrast to pions.
The  K$^+$ yield increases by a factor of about
4 when using an $\alpha$ beam instead of a deuteron beam
whereas the total pion yield increases only by factor of about 2.
According to  transport 
calculations, the K$^+$ mesons are predominantly 
produced in secondary collisions involving pions or/and baryonic resonances.
In addition, the calculations demonstrate that not only the number
of hadron-hadron collisions (which produce K$^+$ mesons) 
increases linearly with the number 
of projectile nucleons, but also 
the average value of the energy available in a hadron-hadron collision.
This effect is able to explain the observed projectile dependence of the
K$^+$ yield.

This work is supported by the German Federal Government and
by the Polish Committee of Scientific Research (contract no. 2P03B 111 09).

$^{*}$ Present address: University of Bern, 3012 Bern, Switzerland

$^{\dagger}$ Present address: SUBATECH, IN2P3-CNRS, F-44070 Nantes, France

$^{\ddagger}$ Present address: 
Institut de Physique Nucl\'{e}aire, IN2P3-CNRS, F-91406 Orsay, France

~\\

\newpage

{\bf Fig.1}:
Inclusive $\pi^+$ production cross sections $d^{3}\sigma / dp^{3}$ vs kinetic
energy of pions in the nucleon-nucleon center of mass frame measured 
at $\theta _{lab} = 40^{\rm o}$.
Left side: d+C and d+Pb collisions, right side: $\alpha$+C and $\alpha$+Pb 
collisions.
The lines represent Maxwell-Boltzmann distributions fitted to the data
(cf. Table~I).

{\bf Fig.2}:
Upper part: ratio of pion differential cross sections as a function of the pion
kinetic energy in the center of mass frame measured for the same projectile
(d or $\alpha$) on different targets (Pb and C). The solid line is the
corresponding ratio for a proton projectile [13].
Lower part: pion ratios measured for the same target nucleus (Pb or C) but
for different projectiles ($\alpha$ and d). The solid line represents
the $\pi^+$
ratio for d+Pb/p+Pb. The arrow denotes the maximum pion energy which can be
obtained in free N-N collisions at a bombarding energy 1.15~GeV/nucleon.

{\bf Fig.3}:
Inclusive K$^+$ cross sections $d^{3}\sigma / dp^{3}$ vs kinetic
energy of kaons in the nucleon-nucleon center of mass frame measured 
at $\theta _{lab} = 40^{\rm o}$.
Upper part: d+C and d+Pb collisions,
lower part: $\alpha$+C and $\alpha$+Pb collisions.
The lines are Maxwell-Boltzmann fits to the data (cf. Table~II).

{\bf Fig.4}:
BUU calculation of the 
double differential K$^+$ production cross sections  
(at $\theta_{lab}=40^o$, E$_{beam}$=1.15 GeV/nucleon) 
as a function of the laboratory
momentum for collisions of d+C (upper left panel), $\alpha$+C (upper right
panel), d+Pb (lower left panel) and $\alpha$+Pb (lower right panel) in 
comparison with the data (full symbols). The solid lines correspond to the 
total K$^+$ yield. The dotted, dashed and dashed-dotted lines represent
contributions of pion-baryon, nucleon-resonance and nucleon-nucleon collisions,
respectively.

\begin{table}
{\centering TABLE I.
Fit parameters and resulting cross sections for $\pi^{+}$ production.
The numbers in brackets give a one-$\sigma$ uncertainty
of the last digits.}
\begin{tabular}{ccccc}
&A$_0$(b/(GeV/c)$^3$)&T$_{0}$ (MeV) & $d\sigma/d\Omega$ (mb/sr) & $\chi^{2}$/NDF\\
\hline
d+C\phantom{b}         & \phantom{2}9.3(4) & 46(3) &
 \phantom{1}10.1(4)           & 1.40 \\
$\alpha$+C\phantom{b}  & \phantom{2}6.2(3) & 55(5) &
 \phantom{1}9.6(6)           & 0.87 \\
d+Pb        & 29(1) & 48(3) &           36(1) & 1.17 \\
$\alpha$+Pb & 25(1) & 57(5) &           43(3) & 0.97 \\
\end{tabular}
\end{table}
\begin{table}
{\centering TABLE II.
Fit parameters and resulting cross sections for K$^+$ production.
The numbers in brackets give a one-$\sigma$ uncertainty of the last digits.
The T$_{0}$ parameters for C marked with $^*$ were not fitted but
adopted from the Pb data.}
\begin{tabular}{ccccc}
&A$_0$(mb/(GeV/c)$^3$)&T$_{0}$ (MeV)&$d\sigma/d\Omega$ ($\mu$b/sr)&$\chi^{2}$/NDF\\
\hline
d+C\phantom{b}         & 0.14(3)           & 31$^*$ &
 0.4(1)            & 0.43 \\
$\alpha$+C\phantom{b}  & 0.36(6)           & 38$^*$ &
 1.4\phantom{0}(2)  & 0.70 \\
d+Pb        & 1.8(2) & 31(8) & 4.7(6)  & 1.48 \\
$\alpha$+Pb & 5.0(5) & 38(4) & 20(2) & 0.72 \\
\end{tabular}
\end{table}

\mbox{\epsfig{file=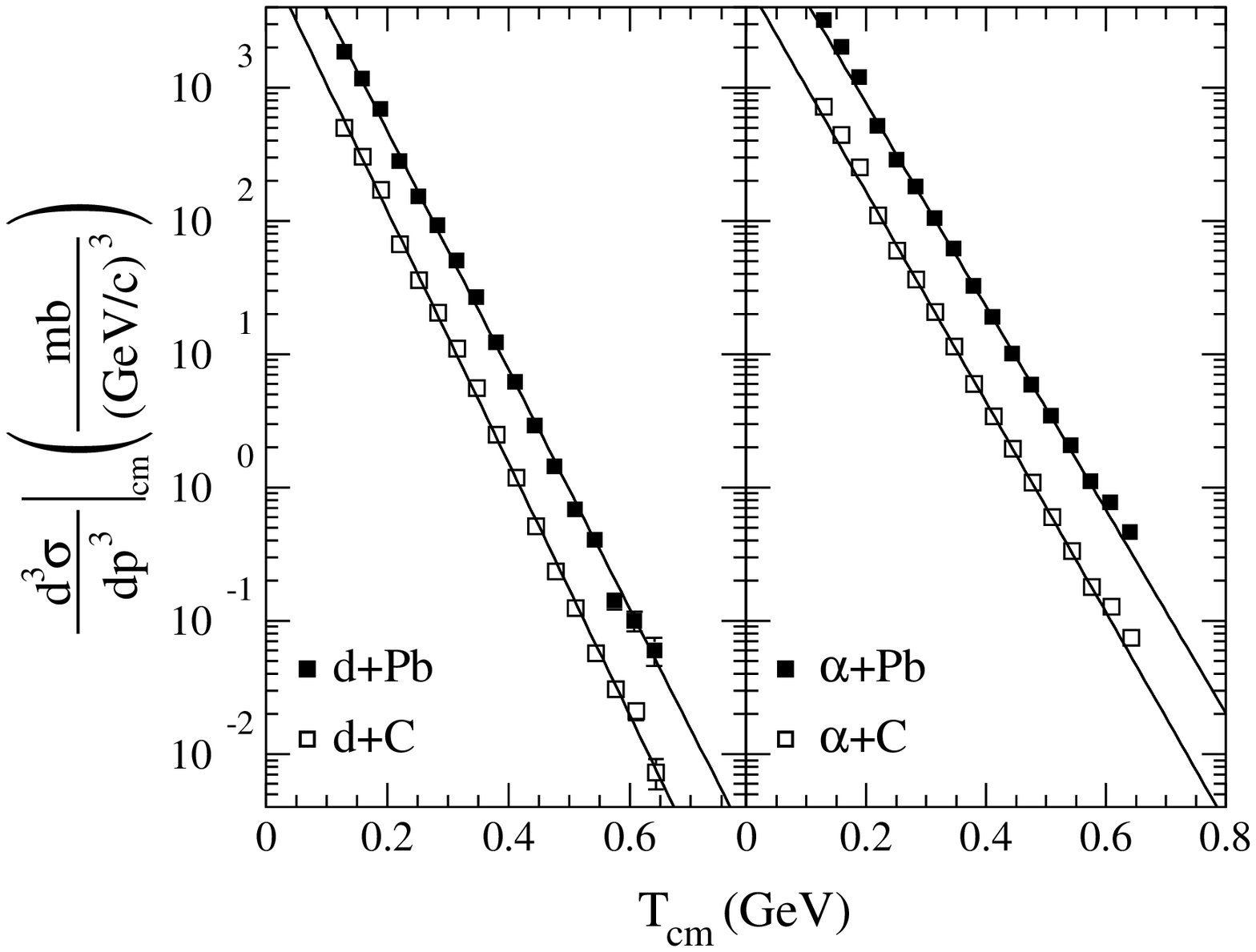,width=16.cm}}
                          
\mbox{\epsfig{file=figure2.ps,width=12.cm}}
                          
\mbox{\epsfig{file=figure3.ps,width=12.cm}}
                          
\mbox{\epsfig{file=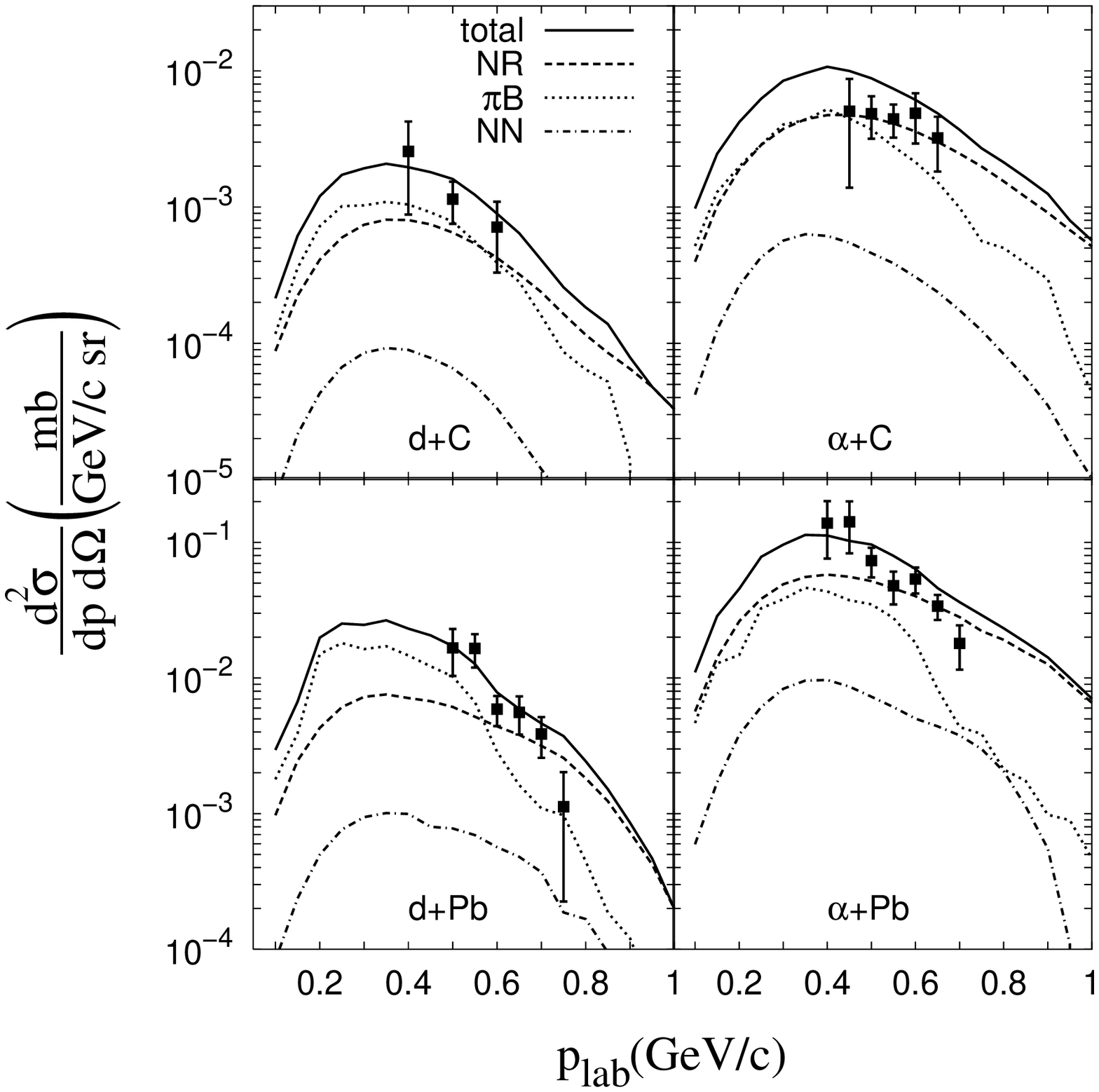,width=16.cm}}

\end{document}